**Model-agnostic Knowledge Augmented Machine Learning for Weakly Annotated Bioimaging Data**

*Jayanie Bogahawatte[1], Sachith Seneviratne[1], Simon Maksour[2,3], Dzung Do-Ha[2], Elizabeth Qian[4],* Samuel Brennan[5], Matthew Keon[5,6], *Lezanne Ooi[2*], Saman Halgamuge[1*]*

[1] *AI, Optimization and Pattern Recognition Research Group, Faculty of Engineering and Information Technology, The University of Melbourne, Melbourne, Victoria, Australia*

[2] School of Science and Molecular Horizons, *University of Wollongong, Northfields Avenue, New South Wales, Australia*

[3] *Charles Perkins Centre, The University of Sydney, Sydney, New South Wales, Australia*

[4] *The Florey Institute of Neuroscience and Mental Health, The University of Melbourne, Melbourne, Victoria, Australia*

[5] *GenieUs Genomics Pty Ltd, Sydney, NSW, Australia*

[6] *Xstendyl, Sydney, NSW, Australia*

**Corresponding Authors: saman@unimelb.edu.au, lezanne@uow.edu.au*

# Abstract

Machine learning (ML) models are increasingly used for automated feature extraction from weakly annotated bioimaging data, such as microscopy and cytomorphology images. Due to the sole dependency of automated feature extraction, existing methods yield inconsistent classification performance across weakly annotated datasets. Here, we propose a model-agnostic knowledge augmentation method that enables ML models to exploit domain knowledge beyond what automated feature extraction can capture. The proposed approach improves classification performance across seven ML architectures. When integrated with masked autoencoder-based self-supervised learning, our method achieves the strongest overall performance, improving precision by 2.9% over the next-best method on three in-house microscopy datasets and by an average of 6.8% across three public cytomorphology benchmarks. We perform nucleus-level classification based on mislocalization of TAR DNA-binding protein 43 (TDP-43) from the nucleus to the cytoplasm, a phenotype linked to Amyotrophic Lateral Sclerosis (ALS), using three independent microscopy datasets of motor neurons derived from human induced pluripotent stem cells. Accurate quantification of this phenotype at scale remains challenging and may contribute to inconsistencies across studies. Application of the best-performing model to these microscopy datasets reveals biological interpretations related to ALS, including dose-dependent effects of the proteasomal stressor, MG132, and the oxidative stressor, sodium arsenite, on TDP-43 mislocalization, variable pathology in motor neurons from sporadic and familial ALS donors, and a lack of protection of the pharmacological compounds, edaravone and niclosamide, against sodium arsenite-induced TDP-43 mislocalization. Together, our pipeline enables in-vitro cellular image analysis to advance understanding of ALS-associated pathology and potential treatments.

# Introduction

Motor Neuron Diseases comprise a group of progressive neurodegenerative disorders characterized by the degeneration of motor neurons, leading to the gradual loss of voluntary muscle control. Among them, Amyotrophic Lateral Sclerosis (ALS) is a prevalent disease, involving both upper and lower motor neurons and resulting in progressive muscle weakness, paralysis, and ultimately respiratory failure [1]. ALS can be broadly classified into sporadic (sALS) and familial (fALS) forms, with sALS accounting for approximately 90-95% of all cases and fALS representing the remaining 5-10% [2]. While the genetic causes of fALS have been extensively characterized, the underlying genetic and molecular drivers of sALS are less well established. Moreover, disease characteristics often exhibit substantial inter-individual variability, adding further complexity to modelling disease mechanisms [3]. Cytoplasmic mislocalization and aggregation of TAR DNA-binding protein 43 (TDP-43) within motor neurons is a defining disease characteristic observed in approximately 97% of all ALS cases [4]. The depletion of nuclear TDP-43 and its cytoplasmic accumulation has been linked to widespread molecular dysfunction, including aberrant mRNA splicing and stability, retention of cryptic exons, impaired DNA damage repair, cytoskeletal disorganization, and mitochondrial defects [5,6].

Recent advances in human induced pluripotent stem cell (iPSC) technology have enabled robust in vitro modelling of ALS pathology [7]. The iPSCs can be differentiated into motor neurons, enabling the investigation of disease mechanisms under controlled conditions. These iPSC-derived motor neurons have been used to investigate TDP-43 mislocalization induced by exogenous stressors, such as proteasome inhibition, oxidative stress, and heat stress [8–11], and to understand the differential patterns of TDP-43 mislocalization between donors as a critical step toward clarifying ALS heterogeneity [12]. Additionally, the comprehensive characterization of how drug compounds influence the localization of TDP-43 is essential for delineating disease mechanisms and to enable personalized medicine [12–14]. Therefore, analysis of immunofluorescence microscopy imaging obtained using iPSC-derived motor neurons serves as a crucial approach to quantify and compare TDP-43 mislocalization dynamics.

Many pathological processes are closely associated with the spatial distribution of specific proteins within cells [15]. Accordingly, TDP-43 mislocalization can be formulated as a single-cell subcellular localization problem, where the objective is to classify cells based on the intracellular distribution of TDP-43. Existing computational approaches have primarily employed machine learning models to capture global image-level patterns [16–20], often emphasizing cell segmentation rather than fine-grained single-cell localization [16,21]. While deep neural networks (DNNs) have demonstrated strong performance in cell localization and classification tasks [22–24], their success heavily relies on large-scale annotated datasets. This requirement is particularly challenging in microscopy imaging, where individual images may contain hundreds or thousands of cells and accurate nucleus-level annotation requires labour intensive biological expertise. Even though weakly supervised learning frameworks that use image-level labels to infer cell-level features have been explored [25,26], these approaches are suboptimal for cases like TDP-43 pathology, where varying degrees of mislocalization are observed across hundreds or thousands of cells within each image and across all the images within the datasets. Furthermore, challenges exist in cytomorphology, where cell classification relies on the visual assessment of cellular morphology. In cytomorphological analysis, experts explore features such as nuclear shape and size, chromatin organisation, nucleoli, cytoplasmic appearance, granularity, maturation stage, and the presence of atypical or malignant characteristics [27]. Both protein localization analysis and cytomorphological classification therefore depend on the extraction of biologically meaningful image features at the level of individual cells. Traditionally, such features are manually designed based on domain expertise and prior biological knowledge [28,29], offering interpretability but often failing to capture the full complexity of protein localization at the single-cell level. Deep learning based methods address this limitation by automatically learning feature representations directly from images and have become a dominant approach in biological image analysis [22–24]. However, existing approaches typically rely solely on automatically learned features and cannot explicitly exploit biologically meaningful prior knowledge accumulated through decades of experimental research. Under weak supervision, learned representations alone may also be incomplete, as they are not guaranteed to capture subtle yet task-critical localization cues. These observations suggest that biologically derived descriptors and learned visual representations provide complementary sources of information. However, effective integration of biological prior knowledge with automatically learned representations remains an open challenge in computational bioimaging, particularly in annotation-scarce settings.

Considering these limitations, we propose a model-agnostic knowledge augmentation method for weakly annotated bioimaging data. The proposed framework combines biologically derived prior knowledge with learned visual representations through an adaptive weighting mechanism that automatically determines their relative contributions. Unlike conventional approaches that rely solely on automated feature extraction, the proposed method enables ML models to exploit complementary information from both learned image representations and biologically informed descriptors. We additionally introduce a pipeline that comprises of three main components to analyse the microscopy images. First, a microscopy image pre-processing module extracts images that contain a single nucleus per image, which are later referred to as “Single Nucleus Images (SNIs)” (Refer Figure 1 Module (a)). We demonstrate that this module is adaptable to distinct microscopy image datasets that are acquired in different laboratories under varying experimental settings. Second, a Masked Autoencoder [30] based self-supervised learning module learns feature representations from unlabelled SNIs (Refer Figure 1 Module (b)). Third, the proposed knowledge augmentation module integrates self-supervised representations with biologically derived descriptors through adaptive weighting (Refer Figure 1 Module (c)). Importantly, our knowledge augmentation method is model-agnostic and can be integrated with diverse machine learning architectures. Our key contributions are summarised as follows:

1. We propose a model-agnostic knowledge augmentation method that explicitly integrates prior knowledge with learned visual representations for weakly annotated bioimaging data.
2. We evaluate our knowledge augmentation method on three in house independent microscopy datasets and three publicly available cytomorphology datasets and demonstrate the performance improvements considering seven ML architectures. We also demonstrate that the combination of the proposed knowledge augmentation method with self-supervised learning using MAE model consistently achieves the best performance across the six datasets. In addition, we introduce a microscopy image pre-processing and SNIs generation module for our in house microscopy image datasets.

3. We demonstrate the applicability of our pipeline across key ALS research contexts, including environmental toxin exposure, disease subtype stratification, and drug screening.

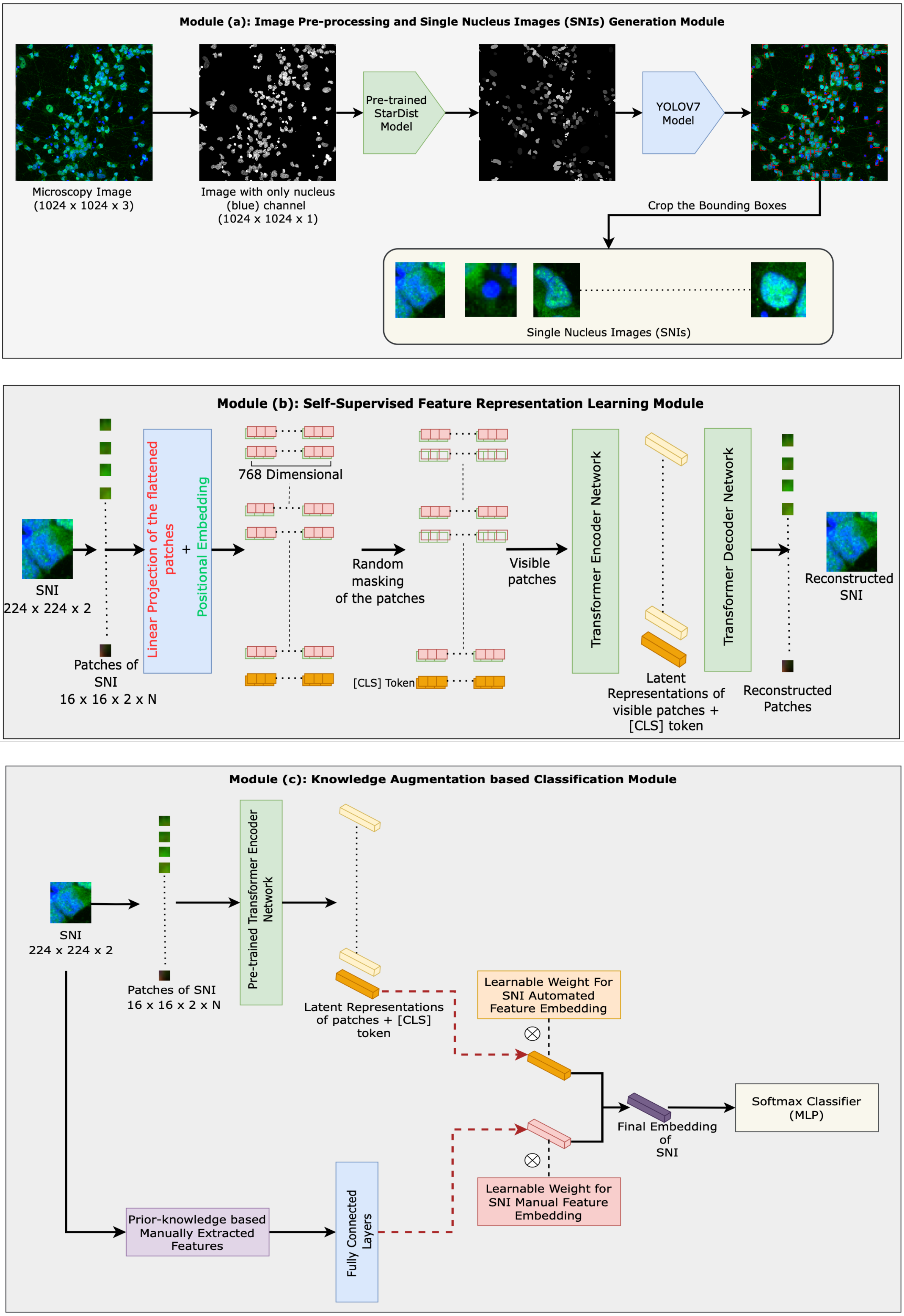
Module (a): Image Pre-processing and Single Nucleus Images (SNIs) Generation Module
Microscopy Image (1024 x 1024 x 3)
Image with only nucleus (blue) channel (1024 x 1024 x 1)
Pre-trained StarDist Model
YOLOV7 Model
Crop the Bounding Boxes
Single Nucleus Images (SNIs)
Module (b): Self-Supervised Feature Representation Learning Module
SNI 224 x 224 x 2
Patches of SNI 16 x 16 x 2 x N
Linear Projection of the flattened patches
+
Positional Embedding
768 Dimensional
Random masking of the patches
[CLS] Token
Visible patches
Transformer Encoder Network
Latent Representations of visible patches + [CLS] token
Transformer Decoder Network
Reconstructed Patches
Reconstructed SNI
Module (c): Knowledge Augmentation based Classification Module
SNI 224 x 224 x 2
Patches of SNI 16 x 16 x 2 x N
Pre-trained Transformer Encoder Network
Latent Representations of patches + [CLS] token
Learnable Weight For SNI Automated Feature Embedding
Final Embedding of SNI
Softmax Classifier (MLP)
Learnable Weight for SNI Manual Feature Embedding
Prior-knowledge based Manually Extracted Features
Fully Connected Layers

Figure 1: Overview of the proposed pipeline. The three modules in the proposed pipeline corresponding to the three main steps described in the Introduction section. Module (a): Microscopy image pre-processing and SNIs generation module. Module (b): Self-Supervised feature representation learning module. Masked Autoencoder (MAE) model [30] is leveraged as the underlying architecture of this module. Module (c): Model-agnostic Knowledge augmentation method-based classification module. This module integrates automated features obtained from any ML model with prior knowledge based features. Here, the pre-trained encoder of the self-supervised feature representation learning module is used for the automated feature extraction.

# Results

## Microscopy Datasets

Mislocalization and aggregation of TDP-43 were examined using immunofluorescence microscopy images of motor neurons differentiated from iPSCs derived from both healthy controls and ALS donors. Our analysis incorporated three in-house independent microscopy image datasets, each acquired under varying experimental settings and using different imaging techniques. The heterogeneity of these datasets strengthens the robustness of our findings and supports the generalisability of the proposed pipeline. The three datasets correspond to complementary applications in ALS research. Dataset 1 was used to assess the impact of cellular stressors on TDP-43 pathology, Dataset 2 was used for ALS disease subtype classification [12], and Dataset 3 was used in a drug discovery context. Across all datasets, the microscopy images were in the size of 1024 x 1024 with two colour channels as blue and green which indicated nucleus (Hoechst stain) and TDP-43 protein (by immunocytochemistry), respectively.

### Microscopy Dataset 1 ($D_1$)

Dataset 1 consisted of microscopy images obtained from spinal motor neurons derived from two healthy control iPSC lines. To induce pathological features associated with ALS, the motor neurons were exposed to two extrinsic stressors with distinct mechanisms of action: a proteasomal inhibitor, Z-Leu-Leu-Leu-al (MG132) and an oxidative stressor, sodium arsenite (SA). Two different paradigms used in the literature include a chronic (16 h application) of MG132 was applied at 1 μM and 10 μM concentrations, and an acute application of high concentrations of SA at 0.1 mM and 0.5 mM concentrations for 30 minutes, representing low and high levels of each stressor. An untreated condition was included as a baseline for comparison to assess dose dependent effects. In total, 72 microscopy image stacks were collected across all experimental conditions. Each stack contained 35 images along the z-axis, thereby capturing the sequential optical sections along the tissue depth. To facilitate downstream analysis using the proposed pipeline, the z-stacks were processed in ImageJ [31], where all slices within a stack were combined by maximum-intensity projection. This procedure generated a single two-dimensional RGB image per stack.

### Microscopy Dataset 2 ($D_2$)

Dataset 2 comprised microscopy images from motor neurons derived from iPSCs of 21 healthy controls, 42 sALS donors, and 5 fALS donors [12]. The fALS cohort included individuals carrying distinct disease-associated genotypes, including SOD1 (G93A and I114T), C9ORF72 expansion, TDP-43 (S393L), and UBQLN2 (T487I). Each image included a z-stack of sequential optical slices, containing 10 slices per stack. For preprocessing, the stacks were permuted in ImageJ [31] to generate a single representative RGB image per stack. Preliminary experiments indicated that maximum-intensity projection yielded images with excessive brightness and contrast, obscuring finer structural details. Therefore, unlike $D_1$, where maximum-intensity projection was employed, here a summed-intensity projection was selected. Summed-intensity projection provided a more balanced representation of the underlying signal by integrating contributions across slices, thereby preserving subtle morphological features essential for downstream computational analysis.

### Microscopy Dataset 3 ($D_3$)

Dataset 3 comprised microscopy images to assess the impact of potential compounds that could protect against mislocalization; these were obtained from ALS-discordant monozygotic twins (healthy control and a sALS donor). Two drug compounds, Edaravone (ED) and Niclosamide (NA), were administered at 10 μM concentration level to assess their potential to reduce TDP-43 mislocalization. To induce pathological features related to TDP-43, the motor neurons were then subjected to SA for 1 h at 500 μM concentration level. An additional untreated condition was included as a baseline control. In total, 960 microscopy image stacks were generated across all experimental groups, with each stack acquired as a 5 × 5 tiled scan across the x-y plane. Each stack consisted of 25 sequential optical slices acquired along the z-axis; however, many slices within each stack were out of focus. To obtain representative images suitable for computational analysis, the stacks were processed in ImageJ [31] where only the most focused slices from each channel were retained and combined. This procedure produced a single RGB composite image per stack, capturing the most informative features while minimizing imaging artefacts.

## Single Nucleus Images (SNIs)

Our pipeline implements a streamlined yet robust approach for segmenting nuclei from microscopy images to extract SNIs (Refer Figure 1 Module (a)). Across the three datasets, SNIs were categorized into two main classes according to the subcellular localization of TDP-43 protein: (i) normal SNIs ($SNI_{norm}$), with physiological nuclear localization of TDP-43, and (ii) SNIs exhibiting TDP-43 mislocalization ($SNI_{mis}$), where TDP-43 had translocated to the outside of the nucleus. In $D_1$, an additional class was defined to account for cellular debris ($SNI_{deb}$). In $D_2$ and $D_3$, subsets of SNIs displayed unusually high green-channel intensity, necessitating their temporary treatment as a separate class ($SNI_{green}$). These SNIs were subsequently reassigned to one of the two primary classes following further analysis based on unsupervised clustering, ensuring consistent classification across datasets (Refer Methods Section). A limited subset of labelled SNIs was curated from each dataset for training and evaluation of the classification module. Across all three datasets, 10 labelled SNIs per class were used for training. For evaluation, 40 labelled SNIs per class were used for $D_2$ and $D_3$, while 36 labelled SNIs per class were used for $D_1$.

## Cytomorphology Datasets

We used three publicly available peripheral blood cytomorphology datasets encompassing normal and pathological leukocyte morphologies acquired at different institutions using distinct imaging platforms. The dataset introduced by Matek et al. [32] (Matek dataset) comprises of 18,365 single cell images of size 400 x 400 pixels. These images are categorized into 13 distinct classes namely: Myeloblast (MYO), Promyelocyte (PMO), Myelocyte (MYB), neutrophil stab granulocyte (NGB), neutrophil segmented granulocyte (NGS), Basophil (BAS), Eosinophil (EOS), Lymphocyte (LYT), Erythroblast (EBO), and Monocyte (MON). The dataset published by Acevedo et al. [33] (Acevedo dataset) comprises 17,092 images of individual peripheral blood cells acquired using a CellaVision DM96 automated digital microscopy system and are provided as 360 x 363 pixel JPEG images. Cells were obtained from individuals without infection, haematological or oncological disease and without pharmacological treatment at the time of blood collection. Expert clinical pathologists annotated the images into eight major morphological groups: neutrophils, eosinophils, basophils, lymphocytes, monocytes, immature granulocytes, erythroblasts and platelets. The immature-granulocyte category includes promyelocytes, myelocytes and metamyelocytes. The MLL23 dataset [34] is a large expert-annotated collection of peripheral blood single-cell images. Cells were initially identified from low-magnification scans and subsequently re-imaged using a 40x objective. The resulting images have a spatial size of 288 x 288 pixels, corresponding to a 25 x 25 μm field of view. Following removal of duplicate images, the released dataset contains 41,621 single-cell images assigned by expert cytomorphologists to 18 morphological classes.

## Performance analysis of existing methods on classifying SNIs based on TDP-43 mislocalization

To evaluate the performance under a label-scarce setting, we trained all the classification models using only 10 labelled SNIs per class from each dataset. Classification performance was evaluated on independent labelled test sets from each dataset using accuracy, precision, and F1-score, as reported in Table 1. For the MAE-based approach, the encoder was first pre-trained using self-supervised learning with the training sets of SNIs in all three datasets and subsequently fine-tuned on the limited labelled SNI data. Based on the results in Table 2, the compared methods exhibit dataset-dependent performance. MAE-based method [30] achieves the strongest performances on $D_1$ and $D_3$, while the ResNeXt-based method [32,35] attains the highest scores on $D_2$. Overall, no single existing method demonstrates consistently strong performance across all three datasets, with performance varying substantially depending on experimental and dataset settings. This demonstrates the need for a more adaptable approach that can generalize effectively to heterogeneous microscopy datasets.

| **Method** | **$D_1$** | | | **$D_2$** | | | **$D_3$** | | |
|---|---|---|---|---|---|---|---|---|---|
| | **Accuracy** | **Precision** | **F1-score** | **Accuracy** | **Precision** | **F1-score** | **Accuracy** | **Precision** | **F1-score** |
| ResNeXt [32,35] | 80.56% | <u>84.00%</u> | <u>80.19%</u> | **95.83%** | **96.09%** | **95.83%** | 90.83% | 92.29% | 90.50% |
| ViT [36] | 74.07% | 83.46% | 74.11% | 94.17% | 94.39% | 94.07% | 92.50% | 92.63% | 92.38% |
| EfficientNet [37,38] | 73.15% | 74.35% | 72.30% | 76.67% | 81.27% | 74.45% | 89.17% | 90.76% | 89.33% |
| SwinT [39] | 74.07% | 79.56% | 73.97% | 94.17% | 94.17% | 94.10% | 84.17% | 83.84% | 83.94% |
| HCPL [25] | 81.04% | 81.15% | 79.94% | 88.33% | 90.04% | 88.18% | 80.83% | 81.68% | 81.08% |
| NCA [40] | 78.70% | 82.43% | 76.61% | 86.67% | 86.54% | 86.54% | 85.83% | 85.65% | 85.72% |
| MAE [30] | **84.26%** | **85.43%** | **83.41%** | 94.67% | 94.80% | 94.66% | **94.67%** | **94.86%** | **94.63%** |

Table 1: Classification performance on three datasets in terms of accuracy, precision, and F1-score for each model. The best results are highlighted in bold and second best results are underlined.

## Performance analysis of integrating the proposed knowledge augmentation method

We integrated prior knowledge based features which were obtained from SNIs into the existing classification models using the proposed knowledge augmentation method. As shown in Table 2, incorporating prior knowledge based features with our knowledge augmentation method consistently improved the classification performance in all the methods across all three datasets. In addition, this demonstrates the model agnostic nature of the proposed method and its effectiveness across diverse architectural backbones. The most pronounced improvements were observed for the MAE-based method, when the pre-trained encoder is finetuned jointly with knowledge augmentation. It achieves the best overall performance, outperforming the second-best method in precision by 4.2%, 1.3%, and 3.0% on datasets 1,2, and 3 respectively. The strong and stable performance across all datasets highlights its robustness and generalizability under distinct experimental and imaging settings. These results further indicate that feature representations learned through self-supervised learning can be effectively augmented by incorporating prior knowledge based features via the proposed knowledge augmentation, leading to consistent performance improvements. Based on this robust performance, we designate the MAE model integrated with knowledge augmentation as our proposed method for subsequent analyses.

| **Method** | **$D_1$** | | | **$D_2$** | | | **$D_3$** | | |
|---|---|---|---|---|---|---|---|---|---|
| | **Accuracy** | **Precision** | **F1-score** | **Accuracy** | **Precision** | **F1-score** | **Accuracy** | **Precision** | **F1-score** |
| ResNeXt [32,35] | 80.56% | 84.00% | 80.19% | 95.83% | 96.09% | 95.83% | 90.83% | 92.29% | 90.50% |
| ResNeXt + KA | 82.41% | 86.16% | 81.86% | 95.83% | 96.30% | 95.86% | 91.00% | 92.77% | 91.19% |
| ViT [36] | 74.07% | 83.46% | 74.11% | 94.17% | 94.39% | 94.07% | 92.50% | 92.63% | 92.38% |
| ViT + KA | 78.70% | 85.17% | 77.60% | <u>96.33%</u> | <u>96.37%</u> | <u>96.32%</u> | <u>92.50%</u> | <u>93.27%</u> | <u>92.23%</u> |
| EfficientNet [37,38] | 73.15% | 74.35% | 72.30% | 76.67% | 81.27% | 74.45% | 89.17% | 90.76% | 89.33% |
| EfficientNet + KA | 82.41% | 82.39% | 81.98% | 81.67% | 82.28% | 81.68% | 90.83% | 91.90% | 90.94% |
| SwinT [39] | 74.07% | 79.56% | 73.97% | 94.17% | 94.17% | 94.10% | 84.17% | 83.84% | 83.94% |
| SwinT + KA | 79.63% | 82.18% | 79.95% | 95.00% | 95.07% | 94.97% | 85.83% | 85.97% | 85.27% |
| HCPL [25] | 81.04% | 81.15% | 79.94% | 88.33% | 90.04% | 88.18% | 80.83% | 81.68% | 81.08% |
| HCPL + KA | <u>84.26%</u> | 84.30% | <u>84.07%</u> | 90.00% | 90.89% | 90.02% | 85.00% | 84.98% | 84.96% |
| NCA [40] | 78.70% | 82.43% | 76.61% | 86.67% | 86.54% | 86.54% | 85.83% | 85.65% | 85.72% |
| NCA + KA | 81.48% | 83.00% | 81.09% | 89.17% | 89.27% | 89.21% | 90.83% | 92.24% | 91.01% |
| MAE [30] | 84.26% | 85.43% | 83.41% | 94.67% | 94.80% | 94.66% | 94.67% | 94.86% | 94.63% |
| **MAE + KA (Our method)** | **87.96% (+4.4%)** | **89.79% (+4.2%)** | **87.82% (+4.5%)** | **97.50% (+1.2%)** | **97.67% (+1.3%)** | **97.50% (+1.2%)** | **95.83% (+3.6%)** | **96.09% (+3.0%)** | **95.76% (+3.7%)** |

Table 2: Classification performance on the three datasets in terms of accuracy, precision, and F1-score for each model. Each model has a variation which incorporates the proposed knowledge augmentation with prior knowledge based features. It is indicated by '+ KA' and the rows coloured in blue. The best results are highlighted in bold and second best results are underlined. The performance improvement compared to the second best method is included with the + within brackets.

## Performance analysis of different variations of our method on classifying SNIs based on TDP-43 mislocalization

Since the evaluation results reported in Table 2 indicate that the MAE model integrated with the proposed knowledge augmentation outperforms existing methods, we further conducted ablation studies to systematically evaluate the contribution of each component of the proposed method. As shown in Table 3, we consider four different variations of our classification module. As the initial variation, we finetuned the encoder of the MAE model and employed the resulting feature representations for the SNI classification. The second variation relied solely on the features extracted based on domain knowledge. However, this approach yielded comparatively limited performance, demonstrating that the handcrafted features alone may be unable to capture the relevant morphological features in the SNIs associated with TDP-43 mislocalization. We next combined MAE-derived representations with prior knowledge based features through direct feature concatenation. Integrating domain knowledge in this manner improved the average precision across datasets by 0.7% relative to using MAE features alone, demonstrating the complementary nature of the two feature types. Finally, we introduced a learnable weighted knowledge augmentation mechanism that adaptively balances the contributions of automated and knowledge-driven features. This approach achieved the best performance across all datasets, yielding a further 2.3% improvement in average precision compared with standard concatenation. These results highlight that domain knowledge provides information beyond that captured by automated feature extraction and that optimising the integration of both feature sources is critical for robust classification of TDP-43 mislocalization.

| **Method** | **$D_1$** | | | **$D_2$** | | | **$D_3$** | | |
|---|---|---|---|---|---|---|---|---|---|
| | **Accuracy** | **Precision** | **F1-score** | **Accuracy** | **Precision** | **F1-score** | **Accuracy** | **Precision** | **F1-score** |
| MAE finetuned features | 84.26% | 85.43% | 83.41% | 94.67% | 94.80% | 94.66% | 94.67% | 94.86% | 94.63% |
| Prior knowledge based features | 79.63% | 80.29% | 79.74% | 74.17% | 76.50% | 73.45% | 78.03% | 80.65% | 78.83% |
| MAE finetuned features + Prior knowledge based features weighted equally | 85.19% | 86.75% | 85.15% | 94.17% | 95.04% | 94.12% | 95.00% | 95.35% | 94.89% |
| MAE finetuned features + Prior knowledge based features with proposed knowledge augmentation | **87.96%** | **89.79%** | **87.82%** | **97.50%** | **97.67%** | **97.50%** | **95.83%** | **96.09%** | **95.76%** |

Table 3: Ablation analysis of the proposed knowledge augmentation method for TDP-43 mislocalization classification across three microscopy datasets. Classification performance is reported in terms of accuracy, precision and F1-score. Four variants were evaluated: (1) MAE fine-tuned features only; (2) manually extracted prior knowledge based features only; (3) MAE fine-tuned features combined with manual features through direct concatenation; and (4) MAE fine-tuned features combined with manual features using the proposed weighted knowledge augmentation mechanism. The highest performance for each metric is highlighted in bold.

## Biological interpretation of pathology of TDP-43 mislocalization in ALS

Using the proposed pipeline, we first quantified the TDP-43 mislocalization at the single microscopic image level and then computed either condition-wise or cohort-wise mean mislocalization percentages for comparative analysis. Across the three distinct datasets, we systematically analysed the effects of various stressors and drug compounds on TDP-43 mislocalization, as well as pathological characteristics associated with TDP-43 mislocalization in sALS and fALS patients. $D_1$ was used to assess the impact of MG132 and SA stressors on TDP-43 mislocalization, $D_2$ to investigate TDP-43 mislocalization patterns in motor neurons derived from sALS and fALS donors, and $D_3$ to evaluate the effects of drug compounds: ED and NA on TDP-43 mislocalization (Refer to Methods Section for further details about the datasets).

### $D_1$: Proteasomal stress induces pronounced TDP-43 mislocalization

$D_1$ was used to investigate the effects of proteasomal and oxidative stress on TDP-43 localization in healthy iPSC-derived motor neurons. Motor neurons from two healthy control donors were exposed either to the proteasome inhibitor MG132 at two concentrations for 16 hours or to the oxidative stressor SA for 30 minutes, and TDP-43 mislocalization was quantified using the proposed pipeline. Nuclei were classified into three categories based on TDP-43 localization as $SNI_{norm}$, $SNI_{mis}$, and $SNI_{deb}$ (Refer to Results section for further details). For MG132-treated samples, representative microscopy images with class-labelled bounding boxes are shown in Figure 2a, with corresponding per-image class proportions summarised in Figure 2b. MG132 treatment led to marked increases in both TDP-43 mislocalized nuclei and nuclear debris across representative images. At the donor level, exposure to 1 μM MG132 resulted in a substantial elevation in TDP-43 mislocalization (Figure 2c). In healthy control 1, the proportion of mislocalized nuclei increased from 13.04% under vehicle conditions to 52.61% following 1 μM MG132 treatment ($p<0.01$). MG132 exposure also produced a clear dose-dependent increase in nuclear debris as demonstrated in Figure 2d, with debris levels rising to 53.73% at 10 μM MG132 in healthy control 1 compared to DMSO ($p<0.001$). In contrast, the acute SA treatment elicited modest effects on TDP-43 mislocalization and cell viability under the conditions tested. Representative images and corresponding per-image class proportions are shown in Figure 3a and Figure 3b. In two healthy controls, the proportion of mislocalized nuclei increased slightly from 10.84% under control conditions to 23.26% following high (0.5 mM) SA exposure (Figure 3c). SA-induced changes in nuclear debris were limited (Figure 3d), with minimal variation observed in both healthy controls. Collectively, these findings demonstrate that proteasomal stress induces robust, dose-dependent increases in TDP-43 mislocalization and nuclear degeneration, while oxidative stress exerts a weaker effect under the conditions tested, highlighting that both acute and chronic treatments of stressors can be assessed for TDP-43 mislocalization using this pipeline.

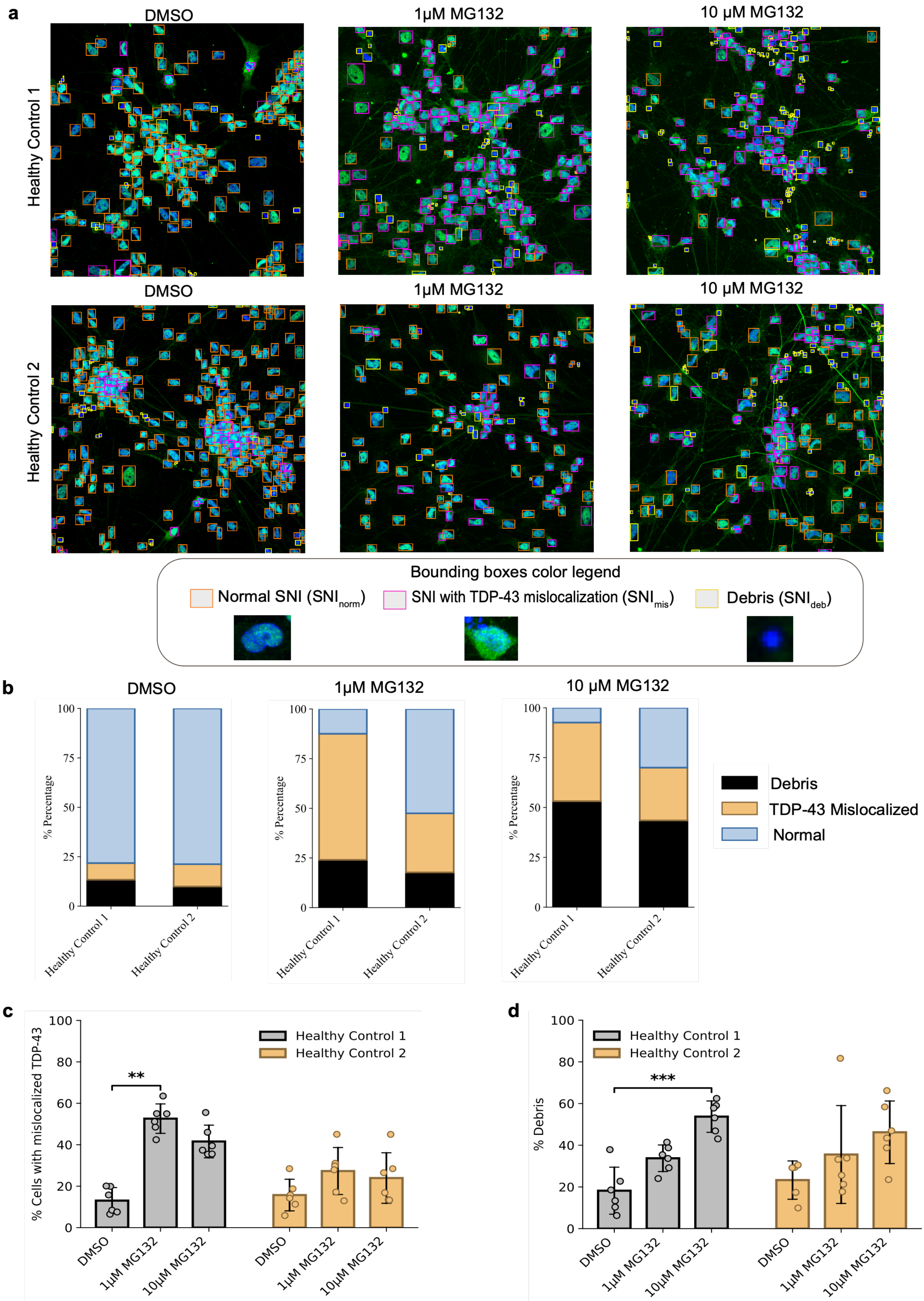

a
DMSO
1μM MG132
10 μM MG132
Healthy Control 1
Healthy Control 2
Bounding boxes color legend
Normal SNI (SNI_norm)
SNI with TDP-43 mislocalization (SNI_mis)
Debris (SNI_deb)
b
DMSO
1μM MG132
10 μM MG132
% Percentage
Healthy Control 1
Healthy Control 2
Debris
TDP-43 Mislocalized
Normal
c
% Cells with mislocalized TDP-43
Healthy Control 1
Healthy Control 2
**
DMSO
1μM MG132
10μM MG132
d
% Debris
Healthy Control 1
Healthy Control 2
***
DMSO
1μM MG132
10μM MG132

Figure 2: Effect of MG132 on the TDP-43 mislocalization on healthy iPSC derived motor neurons. (a) Representative microscopy images with the predicted bounding boxes belonging to the three classes. (b) Representative bar plots corresponding to the microscopy images in (a). Each bar indicates the percentage of nuclei belonging to a specific class. (c) Bar plots representing average percentages of nuclei with TDP-43 mislocalization. Each bar indicates a specific condition including vehicle control, 1 μM MG132, and 10 μM MG132. Each data point represents a microscopy image in the dataset, which was generated from 2 independent differentiations and 3 technical replicates. (d) Bar plots representing average percentages of nuclei debris. Each bar indicates a specific condition including vehicle control, 1 μM MG132, and 10 μM MG132. Each data point represents a microscopy image in the dataset, which was generated from 2 independent differentiations and 3 technical replicates. Statistical significance was assessed using Dunn's post-hoc test with Bonferroni correction, with $* P \leq 0.05$, $** P \leq 0.01$ and $*** P \leq 0.001$.

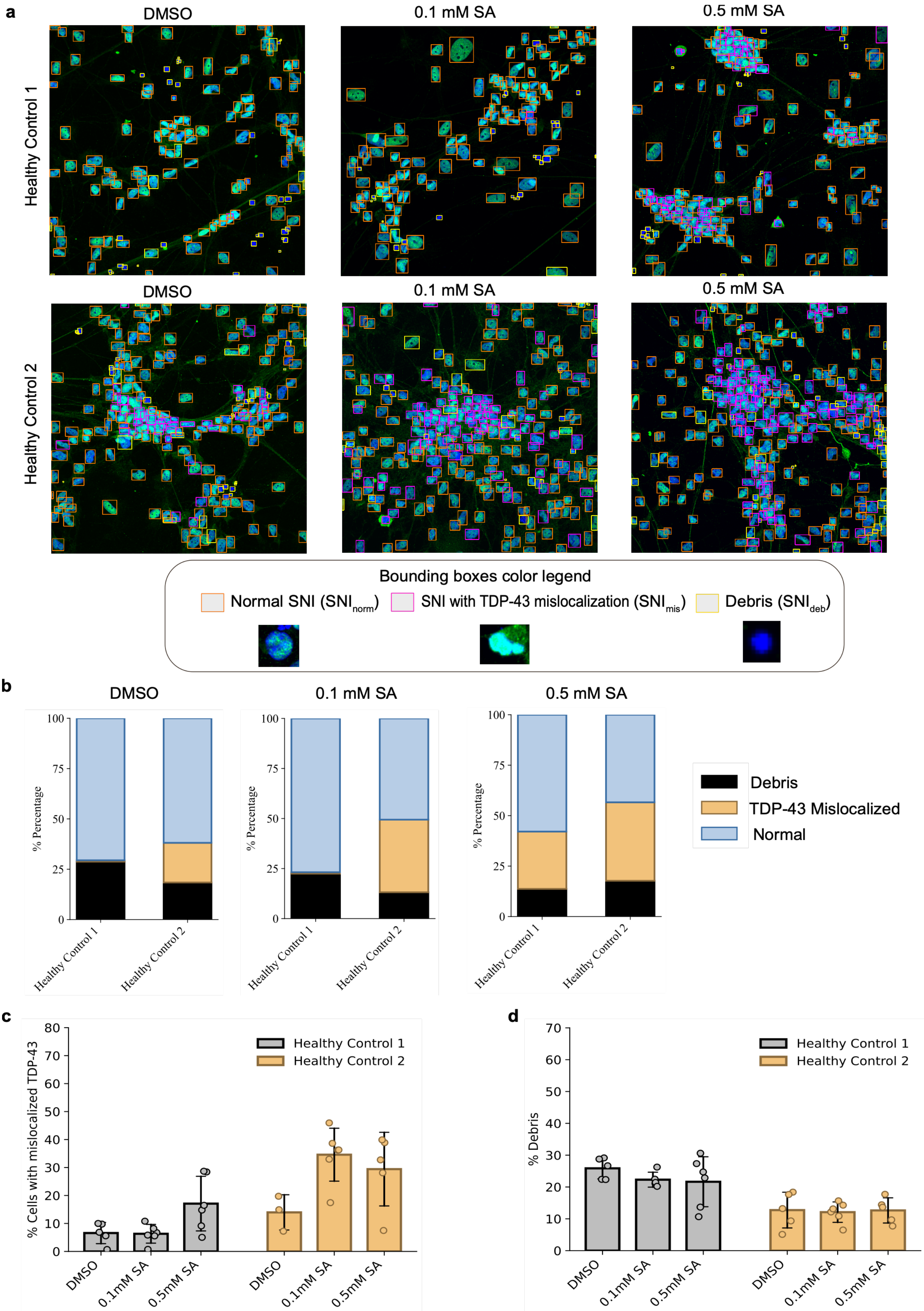
a
DMSO
0.1 mM SA
0.5 mM SA
Healthy Control 1
Healthy Control 2
Bounding boxes color legend
Normal SNI (SNI norm)
SNI with TDP-43 mislocalization (SNI mis)
Debris (SNI deb)
b
DMSO
0.1 mM SA
0.5 mM SA
% Percentage
Healthy Control 1
Healthy Control 2
Debris
TDP-43 Mislocalized
Normal
c
% Cells with mislocalized TDP-43
Healthy Control 1
Healthy Control 2
DMSO
0.1mM SA
0.5mM SA
d
% Debris
Healthy Control 1
Healthy Control 2
DMSO
0.1mM SA
0.5mM SA

Figure 3: Effect of SA on the TDP-43 mislocalization on healthy iPSC derived motor neurons. (a) Representative microscopy images with the predicted bounding boxes belonging to the three classes. (b) Representative bar plots corresponding to the microscopy images in (a). Each bar indicates the percentage of nuclei belonging to a specific class. (c) Bar plots representing average percentages of nuclei with TDP-43 mislocalization. Each bar indicates a specific condition including vehicle control, 0.1 mM SA, and 0.5 mM SA. Each data point represents a microscopy image in the dataset, which was generated from 2 independent differentiations and 3 technical replicates. (d) Bar plots representing average percentages of nuclei debris. Each bar indicates a specific condition including vehicle control, 0.1 mM SA, and 0.5 mM SA. Each data point represents a microscopy image in the dataset, which was generated from 2 independent differentiations and 3 technical replicates.

## $D_2$: Elevated TDP-43 mislocalization in both sporadic and familial ALS

$D_2$ was used to assess disease-associated alterations in TDP-43 localization by comparing motor neurons derived from patients with sALS and fALS, to those from healthy controls. As summarised in Table 4, both ALS cohorts exhibited higher levels of TDP-43 mislocalization relative to healthy controls, consistent with ALS-associated disruption of TDP-43 regulation. Specifically, TDP-43 mislocalization increased by 14.4% in the sALS cohort and by 28.5% in the fALS cohort compared with controls. Analysis of microscopy images from 21 healthy control, 42 sALS, and 5 fALS iPSC-derived motor neuron lines revealed a progressive increase in TDP-43 mislocalization across the three groups (Figure 4c). Healthy control motor neurons showed an average mislocalization of 17.61%, which increased to 20.14% in sALS and further to 22.63% in fALS motor neurons. However, the population showed high variability and the observed increases were not statistically significant between groups. To further characterise inter-donor variability, TDP-43 mislocalization was analysed at the individual cell line level for both sALS and fALS cohorts as presented in Table 4. Among the fALS cohort, the cell line carrying a C9ORF72 expansion exhibited the highest degree of TDP-43 mislocalization, which is represented by fALS cell line 114 in Table 4. In addition, TDP-43 mislocalization levels were also significantly increased in cell lines carrying C9ORF72 and SOD1 mutations, compared with the mean TDP-43 mislocalization percentage of healthy controls ($p \leq 0.05$) as shown in Table 4. This observation aligns with previous reports identifying C9ORF72 mutations as strongly associated with TDP-43 pathology in ALS [41]. To investigate whether TDP-43 mislocalization was associated with the motor-neuron degeneration phenotype previously identified in patient-derived sALS lines, sALS donor lines exhibiting reduced LD50 survival (LD50 < 47 days) [12] were analysed individually and the results are presented in Table 4. LD50 represents the time at which total motor neuron neurite length declines to 50% of its maximum and therefore provides a measure of accelerated motor neuron degeneration [12]. TDP-43 mislocalization percentage for each phenotype positive sALS line was compared with the mean TDP-43 mislocalization percentage of healthy controls. Of the 18 phenotype-positive sALS lines analysed, 3 lines (IDs: 110, 121, 106) exhibited significantly elevated TDP-43 mislocalization compared with healthy controls ($p \leq 0.05$). To determine whether these differences were specific to sALS lines exhibiting the reduced-survival phenotype, the significantly altered lines were subsequently compared with sALS lines not classified as phenotype-positive. All the 3 sALS lines also exhibited significantly higher TDP-43 mislocalization than the non-phenotype sALS group ($p \leq 0.05$). These findings indicate that increased TDP-43 mislocalization is present in a subset of sALS donor lines exhibiting accelerated motor-neuron degeneration and is not uniformly observed across sALS lines. Overall, these results support the view that aberrant TDP-43 localization is a common hallmark of ALS, irrespective of genetic aetiology [4].

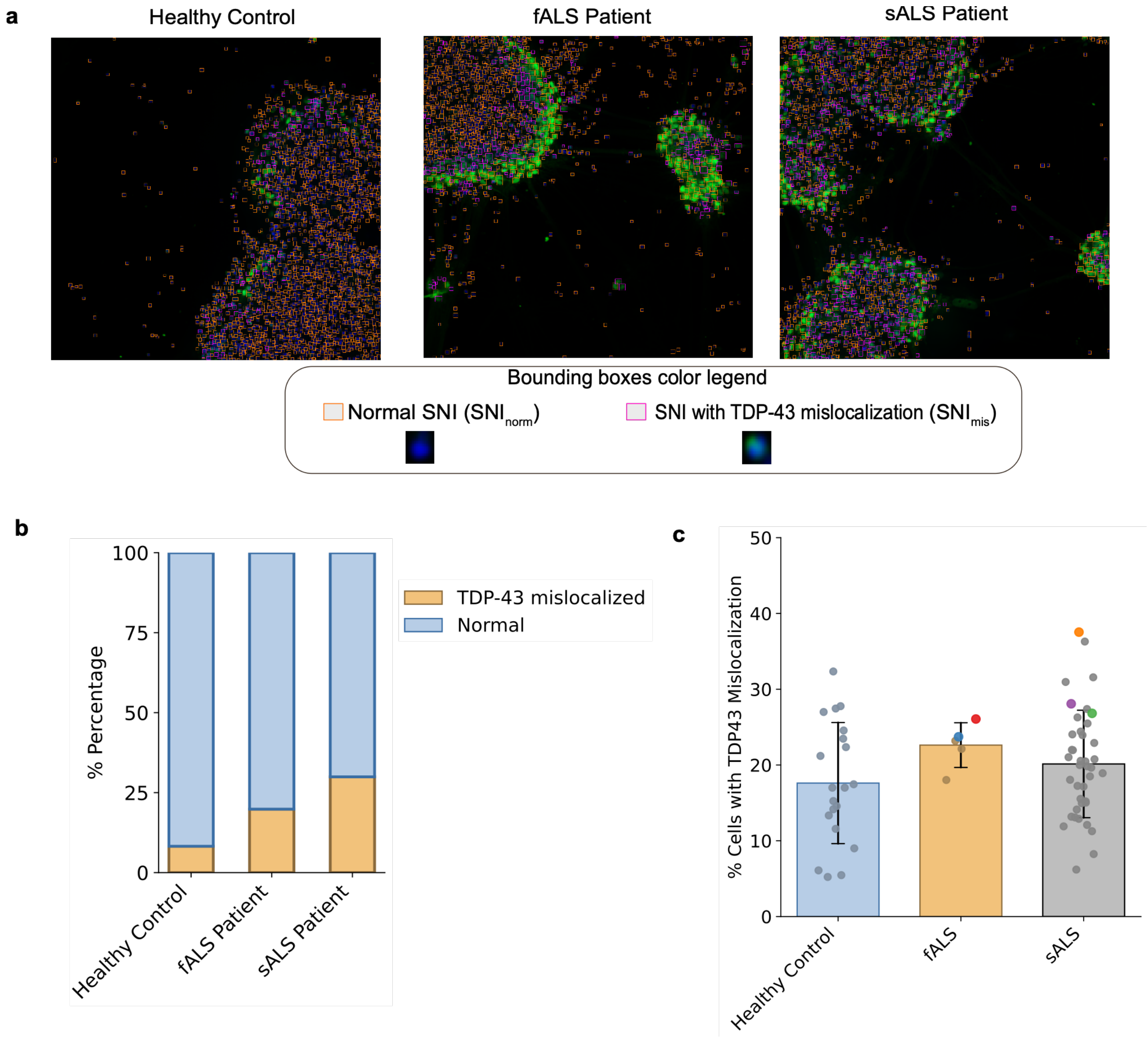

Figure 4: TDP-43 mislocalization analysis on healthy, sALS, and fALS iPSC derived motor neurons. (a) Representative microscopy images with the predicted bounding boxes belonging to the two classes. (b) Representative bar plots corresponding to the microscopy images in (a). Each bar indicates the percentage of nuclei belonging to a specific class. (c) Bar plots representing average percentages of nuclei with TDP-43 mislocalization in healthy, sALS and fALS cell lines. Each data point represents a cell line. There are 21 healthy controls, 42 sALS patients, and 5 fALS patients. Each cell line with a significantly elevated TDP-43 mislocalization percentage compared to the mean TDP-43 mislocalization percentage of the healthy control cohort is indicated by distinct coloured data points (fALS cell line 114: red, fALS cell line 84: blue, sALS cell line 110: green, sALS cell line 121: purple, sALS cell line 106: orange).

| Cell Line ID | TDP-43 mislocalization % | p-value compared to the healthy control cohort |
|---|---|---|
| sALS cell line 110 | 26.82% | 0.003 |
| sALS cell line 121 | 28.08% | 0.002 |
| sALS cell line 106 | 37.54% | 0.0002 |
| fALS cell line 114 | 26.07% | 0.01 |
| fALS cell line 84 | 23.73% | 0.05 |

Table 4: sALS and fALS cell lines in $D_2$ that have $p \leq 0.05$ compared to the mean TDP-43 mislocalization percentage of the healthy control cohort. TDP-43 mislocalization percentage of each cell line is also reported. fALS cell line 114 and 84 carry C9ORF72 and SOD1 mutations, respectively.

## $D_3$: Oxidative stress amplifies disease-associated TDP-43 mislocalization and reveals differential drug responses

The most comprehensive comparisons were derived from $D_3$ which includes ALS-discordant monozygotic twins (sALS vs healthy control). Motor neurons were treated with oxidative stress inducer SA (500 µM, 1 h) and either ED or NA. Additionally, a combination of both compounds (10 µM, 24 h) was used. Microscopy images acquired following treatment were processed using our proposed analysis pipeline to quantify changes in TDP-43 mislocalization. As demonstrated in Figure 5, motor neurons from both the healthy control and the sALS donor exhibited a similar overall pattern of response. SA exposure increased mislocalization from 8.79% to 11.83% in the healthy control and from 9.52% to 14.98% in the sALS patient ($p<0.05$) (Figure 5b). Treatment with ED demonstrated a trend to reduce this effect, reducing mislocalization to 10.64% in the healthy control and to 12.34% in the sALS donor (Figure 5b). However, in this experimental paradigm, the protective effects of ED were not significantly different from the SA treated positive controls in either cell line. Additionally, neither NA alone nor the combination of the two drugs reduced mislocalization in either donor. Together, these findings suggest that while ED may provide a modest protective effect, it did not reach significance against oxidative stress-induced TDP-43 mislocalization in both healthy and sALS motor neurons. In fact, the combination treatment of NA and ED with SA led to significantly increased TDP-43 mislocalization, compared to SA or SA and ED treatment in the healthy donor line (Figure 5c). While NA has been reported to reduce TDP-43 mislocalization in certain contexts [14], it is also known to disrupt mitochondrial function, which may contribute to the observed increase in mislocalization under oxidative stress conditions [42]. Collectively, these findings indicate that SA increased TDP-43 mislocalization, whereas neither ED nor NA reduced mislocalization. Furthermore, co-incubation with both the compounds was associated with a further increase in TDP-43 mislocalization under the experimental conditions tested.

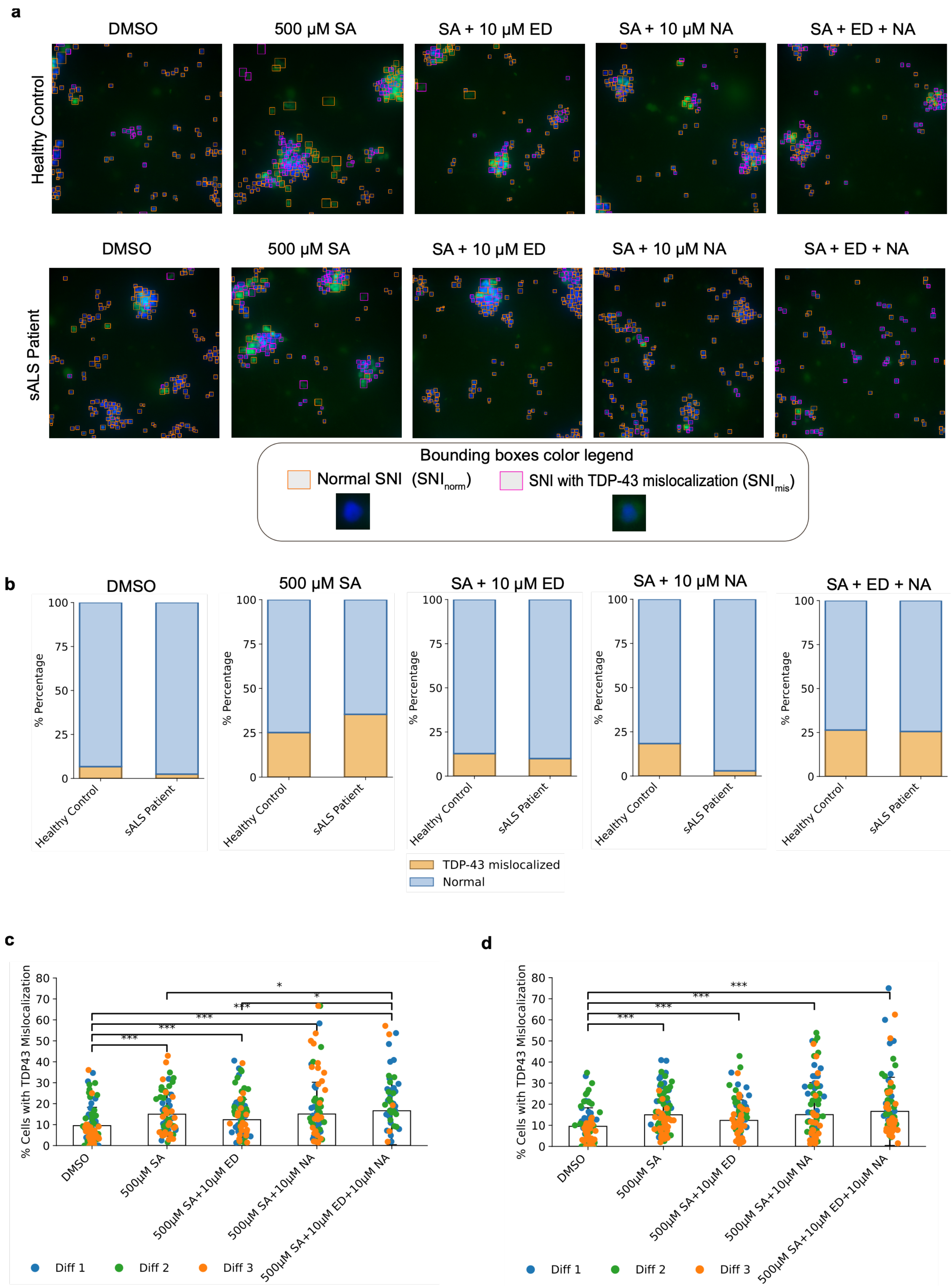

a
DMSO
500 μM SA
SA + 10 μM ED
SA + 10 μM NA
SA + ED + NA
Healthy Control
sALS Patient
Bounding boxes color legend
Normal SNI (SNI_norm)
SNI with TDP-43 mislocalization (SNI_mis)
b
% Percentage
Healthy Control
sALS Patient
TDP-43 mislocalized
Normal
c
d
% Cells with TDP43 Mislocalization
DMSO
500μM SA
500μM SA+10μM ED
500μM SA+10μM NA
500μM SA+10μM ED+10μM NA
Diff 1
Diff 2
Diff 3

Figure 5: Effect of two drug compounds on TDP-43 mislocalization on SA induced healthy and sALS iPSC derived motor neurons. (A) Representative microscopy images with the predicted bounding boxes belonging to the two classes. (B) Representative bar plots corresponding to the microscopy images in (A). Each bar indicates the percentage of nuclei belonging to a specific class. (C) Bar plots representing average percentages of nuclei with TDP-43 mislocalization. Each bar indicates a specific condition including vehicle control, 500 μM SA, 500 μM SA + 10 μM Ed, 500 μM SA + 10 μM NA, and 500 μM SA + 10 μM Ed + 10 μM NA. Each data point represents a microscopy image in the dataset generated from 3 differentiations. Statistical significance was assessed using Dunn's post-hoc test with Bonferroni correction, with $*P \leq 0.05$, $**P \leq 0.01$ and $***P \leq 0.001$.

## Performance analysis of the proposed knowledge augmentation on publicly available cytomorphology bioimaging datasets

To further demonstrate the generalizability and the robustness of the proposed knowledge augmentation method, we conducted experiments using three publicly available cytomorphology datasets of white blood cell (WBC) images: Matek dataset [32], Acevedo dataset [33], and MLL23 dataset [34]. To emulate an annotation-scarce setting, we restricted the training set to 15 labelled images per class in each dataset. Model performance was evaluated on an independent held-out test set for each dataset, reporting the weighted average precision. For the MAE-based model[30], self-supervised pre-training was first performed on the unlabelled training split of Matek dataset to learn generalized visual representations through masked image reconstruction. The pre-trained encoder was then fine-tuned on the downstream WBC image classification task using the limited labelled subset in each dataset. To integrate the prior knowledge based features, we extracted 51 handcrafted features previously shown to be informative for WBC image classification [28]. These included three nucleus shape descriptors, convexity, circularity, and solidity, alongside 48 colour features derived from multiple colour spaces (RGB, HSV, LAB, and YCrCb). Together, these features capture both morphological and chromatic variations relevant to distinguishing WBC subtypes. Given the model-agnostic nature of the proposed knowledge augmentation, we incorporated this mechanism into all the existing classification approaches to assess its effectiveness across diverse model architectures. Therefore, our proposed method integrates automated features from different models with prior knowledge based features via weighted knowledge augmentation.

As shown in Table 5, baseline supervised models trained solely on the limited labelled dataset achieved suboptimal classification performance, reflecting the impact of weak annotation. Incorporating a self-supervised pre-training stage significantly alleviated this constraint, demonstrating the effectiveness of self-supervised representation learning under low-annotation conditions. Once the MAE encoder was pre-trained and subsequently finetuned together with proposed knowledge augmentation, classification performance improved markedly showing an increment of 6.8% in terms of weighted average precision compared to the existing second best method. Additionally, as demonstrated in Table 5, the proposed knowledge augmentation resulted in performance increments in all the existing models. These findings demonstrate that knowledge augmentation with prior knowledge based features provides a robust, model-agnostic enhancement, consistently improving feature representations and downstream performance across diverse architectures.

| Method | Matek Dataset [32] | Acevedo Dataset [33] | MLL23 Dataset [34] |
|---|---|---|---|
| ResNeXt [32,35] | 70.96% | 73.44% | 60.82% |
| ResNeXt + KA | 80.49% | 82.31% | 70.44% |
| ViT [36] | 52.01% | 73.17% | 60.92% |
| ViT + KA | 77.46% | 79.78% | 62.66% |
| EfficientNet [37,38] | 65.42% | 62.24% | 46.71% |
| EfficientNet + KA | 74.48% | 68.49% | 49.52% |
| SwinT [39] | 44.70% | 30.95% | 38.07% |
| SwinT + KA | 67.13% | 37.33% | 58.02% |
| HCPL [25] | 38.51% | 62.65% | 46.33% |
| HCPL + KA | 67.32% | 62.68% | 49.74% |
| NCA [40] | 56.52% | 64.10% | 41.81% |
| NCA + KA | 70.45% | 65.25% | 44.62% |
| MAE [30] | 82.22% | 82.89% | 68.14% |
| MAE + KA (Our method) | 85.67% (+6.4%) | 86.73% (+5.4%) | 76.44% (+8.5%) |

Table 5: Classification performance on cytomorphology datasets in terms of weighted average precision for each model. Our method corresponds to the pipeline with MAE

finetuned features + prior knowledge based features using the proposed knowledge augmentation. Each model has a variation which incorporates the prior knowledge based features using knowledge augmentation, which is indicated by '+ KA' and the rows coloured in blue. The best results are highlighted in bold and second best results are underlined. The performance improvement compared to the second best method is included with the + within brackets.

# Discussion

ALS is characterised by the progressive degeneration of motor neurons, and TDP-43 mislocalization is one of its most prominent pathological hallmarks. However, quantitative assessment of TDP-43 localization at single-cell resolution remains challenging because large-scale expert annotation of microscopy datasets is labour-intensive and often impractical. In this study, we proposed a model-agnostic knowledge augmentation method that enables annotation-efficient classification of weakly labelled bioimaging data by integrating biologically derived descriptors with automatically learned visual representations. Across three independent ALS microscopy datasets and three publicly available cytomorphology datasets, the proposed method consistently improved the performance of diverse machine learning architectures, demonstrating that explicit incorporation of domain knowledge provides complementary information that is not reliably captured through automated feature learning alone.

A central finding of this work is that biological prior knowledge remains highly informative even in the era of self-supervised representation learning. Deep learning models have become the dominant paradigm for bioimage analysis because of their ability to automatically learn complex visual representations directly from raw images. Nevertheless, biological imaging tasks frequently rely on subtle cellular phenotypes that have been established through decades of experimental investigation. For TDP-43 pathology, biologically meaningful attributes such as nuclear morphology, protein intensity distributions and nucleus-cytoplasmic localization ratios provide direct morphological information. Our ablation analyses demonstrate that these manually derived descriptors alone are insufficient for optimal performance, but when combined with self-supervised representations they consistently improve classification accuracy. Importantly, the largest gains were achieved using a learnable weighting strategy rather than simple feature concatenation, suggesting that knowledge integration is most effective when the relative contribution of biological and learned features can be adaptively determined during model optimisation.

The consistent improvements observed across seven machine learning architectures further indicate that the proposed method addresses a limitation that extends beyond any specific model family. Recent advances in bioimage analysis have focused predominantly on developing increasingly powerful architectures, including convolutional neural networks, vision transformers and self-supervised foundation models. Although these approaches differ substantially in their inductive biases and representation learning strategies, all benefited from the proposed knowledge augmentation method. This observation suggests that performance limitations in weakly annotated bioimaging datasets arise not only from model capacity but also from incomplete utilisation of available biological information. By introducing a mechanism that explicitly exploits biologically informed descriptors, our method complements advances in representation learning and offers a broadly applicable strategy for annotation-scarce biological imaging problems.

Beyond methodological advances, the proposed pipeline enabled quantitative investigation of several biological questions relevant to ALS. Across independent experimental settings, we observed induction of TDP-43 mislocalization following proteasomal inhibition (dataset 1), and oxidative stress (dataset 3). Patient-derived motor neurons from both fALS and sALS donors exhibited variability in TDP-43 mislocalization, highlighting the heterogeneous nature of ALS-associated cellular pathology. Furthermore, pharmacological perturbation experiments revealed a lack of protective effects of proposed therapeutic compounds, with ED showing a non-significant trend for mitigating stress-induced pathology whereas NA failed to confer protection under the conditions examined and treatment with both together leading to further increased sodium arsenite-induced TDP-43 mislocalization. Although these findings require validation in larger cohorts and additional experimental systems, they demonstrate that annotation-efficient computational analysis can reveal biologically meaningful phenotypic variation across disease models, environmental perturbations and drug-treatment conditions.

Several avenues for future work emerge from this study. First, the incorporation of richer, domain informed handcrafted features may provide complementary signals that improve classification accuracy. Second, beyond imaging, the proposed pipeline is readily extensible to incorporate additional data modalities, such as patient level clinical reports, where available. In conclusion, we present a model-agnostic knowledge augmentation method for weakly annotated bioimaging data that unifies automatic feature representation learning with biologically informed feature integration. Our results demonstrate that explicitly incorporating prior biological knowledge can consistently enhance machine learning performance across diverse architectures and imaging modalities, particularly under annotation constraints.

# Methods

The proposed pipeline is composed of three main modules as:

1. Image pre-processing and single nucleus images (SNIs) generation module
2. Self-supervised feature representation learning module
3. Model-agnostic knowledge augmentation based classification module

## In-house Microscopy Image Data Collection

### Motor Neuron Differentiation for Dataset 1 and Dataset 3

Motor neurons were differentiated using a modified published protocol [43]. Briefly, motor neurons were differentiated from iPSCs generated from two healthy donors (healthy control 1; HC1 and healthy control 2; HC2, dataset 1) and from ALS-discordant monozygotic twins (sALS *vs* healthy control, dataset 3) as previously described [43]. Briefly, two days before iPSC cultures reached full confluency, iPSCs were transitioned to neuronal induction (NI) medium (DMEM/F12, 1x N2 supplement, 0.2x B27 supplement, 1x non-essential amino acids (NEAA) and 1x GlutaMAX), supplemented with 1 μM LDN193189. After 48 h in NI medium, embryoid bodies (EBs) were generated by lifting the iPSC colonies using dispase and maintaining the cells in suspension with NI medium supplemented with 3 μM CHIR99021 and 2 μM SB431542 for 5 days. EBs were plated down on Matrigel (0.1 mg/mL) coated plates in NI medium supplemented with 3 μM CHIR99021, 2 μM SB431542 and 2 μM DMH-1 for neural rosette formation. After 5 days, neural rosettes were isolated using dispase and plated down on Matrigel-coated plates in motor neuron progenitor media (MNP) (DMEM/F12, 1x N2 supplement, 0.5x B27 supplement, 1x NEAA, 1x GlutaMAX) supplemented with 3 μM CHIR99021, 2 μM SB431542m, 2 μM DMH-1, 0.1 μM retinoic acid and 0.5 μM purmorphamine to caudalise and ventralise the progenitors. MNPs were maintained for 5-7 days with partial media changes every day. During this period, MNPs were gradually transitioned to MNP medium containing BrainPhys instead of DMEM/F12 as basal medium. Once MNP clusters were fully transitioned into BrainPhys-based MNP medium, Accutase was used to dissociate the clusters into single cells. At this point, single cell MNPs were cryopreserved in DMEM/F12 with 10 µM Y-27632, 10% DMSO and 20% Albumax. For final motor neuron (MN) differentiation, MNPs were revived and plated out on Matrigel-coated plates at 80,000 cells/cm$^2$ in MN maturation medium (BrainPhys, 1x N2 supplement, 0.5x B27 supplement, 1x NEAA) supplemented with 0.5 µM retinoic acid, 0.1 µM purmorphamine, 10 µM DAPT, 10 ng/mL BDNF, 10 ng/mL GDNF, 10 ng/mL IGF, 500 ng/mL cAMP and 200 ng/mL L-Ascorbic acid. 10 µM Y-27632 was added for 48 h to promote cell adherence and survival. Partial media changes were performed every 2-3 days. After 10 days, retinoic acid, purmorphamine and DAPT were removed from the medium, and the cells were maintained with MN medium and the remaining supplements for an additional 18 days until experiments were conducted.

### Dataset 1 ($D_1$)

The generated motor neurons were subsequently treated with either MG132, a proteasome inhibitor, or sodium arsenite (SA), an inducer of oxidative stress. Motor neurons were treated with MG132 for 16 hours at 10 µM or 1 µM or with vehicle control (DMSO). SA treatment was added for 30 minutes at 0.1 mM and 0.5 mM and included a no treatment control.

### Dataset 2 ($D_2$)

Cell cultures were washed with phosphate-buffered saline (PBS) and fixed by immersion in 4% (w/v) paraformaldehyde for 2 h at room temperature. Following fixation, samples were processed for immunofluorescence staining. TDP-43 was detected using a rabbit polyclonal anti-TDP-43 antibody (Proteintech, 10782-2-AP; AB_615042) at a dilution of 1:500. Immunolabeling was performed overnight at room temperature in PBS containing 5% normal serum and 0.25% Triton X-100 (Amresco) to facilitate permeabilization and reduce non-specific binding. After primary antibody incubation, samples were washed and incubated with species-appropriate Alexa Fluor-conjugated secondary antibodies (Thermo Fisher Scientific) at a dilution of 1:400 for 2 h at room temperature. All antibody incubations were performed in PBS supplemented with 5% normal serum and 0.25% Triton X-100. Z-stack images were acquired using an Opera Phenix High-Content Screening System (PerkinElmer) at 20× magnification. Four fields of view were captured per well, with four replicate wells analysed per donor.

### Dataset 3 ($D_3$)

Motor neurons were treated with 10 µM of ED (Sigma Aldrich #M70800) or NA (MedChemExpress, #HY-B0497) for 24 h prior to sodium arsenite induced stress. Following drug pre-treatment, a full media change was performed and 500 µM of sodium arsenite was added in motor neuron media for 1 h. Following stress induction, motor neurons were fixed with 4% paraformaldehyde and utilised for immunocytochemistry analysis.

## Image pre-processing and single nucleus images (SNIs) generation module

The initial stage of the proposed pipeline focuses on isolating individual nuclei from whole-field fluorescence microscopy images to enable nucleus-level classification based on the TDP-43 subcellular localization. Acquiring extensive manual annotations of nuclei was not feasible due to the high cellular density within each field of view, often comprising thousands of nuclei per image. Additionally, this task required labour intensive domain-specific expertise to ensure accuracy. Since ground-truth manual annotations including either bounding boxes or pixel-wise segmentation masks were not available for the datasets, supervised object detection or segmentation models could not be directly trained. To address these limitations, we propose a simple, yet innovative pre-processing module as illustrated in Figure 1 Module (a) to extract images which contain single nucleus per image, which are referred to as 'Single Nucleus Images' (SNIs), and to later label a small subset of SNIs for the classification task.

To localize the nuclei in the microscopy images, first the RGB images were split into the three colour channels red, green, and blue. Blue channel indicated the nuclei in the microscopy images. Therefore, the images with only the blue channel were then used as input to a pre-trained nuclei instance segmentation model known as StarDist [44] which was pre-trained on 2D and 3D fluorescence images. Using the pre-trained StarDist model at the inference stage, the nuclei segmentation masks were acquired. Next, the nuclei instance segmentation masks were used to generate coarse bounding boxes. These bounding boxes were represented by the 4 coordinates of the box; xMin, yMin, xMax and yMax which were then used to calculate the centre coordinates, the width and height of the boxes. These values were used to finetune a state-of-the-art object detection model, which was originally pre-trained on natural images, known as YOLOv7 [45]. This step was accommodated to obtain more accurate bounding boxes for the nuclei in microscopy images. The detected nuclear bounding boxes were subsequently expanded by applying a fixed spatial padding margin to each side prior to cropping. This padding was introduced to retain a controlled amount of information outside the nucleus, which is essential for accurately identifying TDP-43 mislocalization. The padded bounding boxes were then cropped to extract the SNIs, ensuring that both nuclear and immediate extra-nuclear regions were preserved for downstream tasks. Further details about the datasets and extracted SNIs are included in Results section.

## Self-supervised feature representation learning module

To enable robust feature representation learning under limited annotated data, we leveraged a self-supervised representation learning strategy. Common self-supervised paradigms include masked image modelling (MIM) [30], contrastive learning [46,47], and generative modelling [48]. Among these distinct approaches, MIM has demonstrated strong performance in medical imaging domains, particularly for X-ray and CT scan analysis [49–52]. Therefore, in this work, we leveraged a MIM-based framework, known as masked autoencoder (MAE) architecture [30], as the self-supervised representation learning module within the proposed pipeline (Refer ure1 Module (b)). The MAE architecture consists of a Vision Transformer (ViT) [36] based encoder and a lightweight transformer decoder. The model is trained to reconstruct masked regions of the input image, thereby encouraging the encoder to learn semantically meaningful latent representations from incomplete visual observations. The asymmetric design, where the encoder processes only visible (unmasked) patches and the decoder reconstructs the full image, improves computational efficiency and scalability during large-scale pretraining. The MAE model was pretrained end-to-end using the extracted SNIs as inputs.

**MAE Encoder:** Each input SNI was resized to 224 × 224 with two channels and partitioned into non-overlapping patches of size 16 × 16. A random subset of patches was masked during training. Consistent with prior work [30], a masking ratio of 75% was applied, which had been shown to enhance representation quality by increasing task difficulty and encouraging global contextual reasoning. Each patch was flattened and linearly projected into a D-dimensional embedding space (D = 768). Learnable positional embeddings were added to preserve spatial information. In contrast to the standard ViT architecture [36], which processes all patch tokens, the MAE encoder operates exclusively on the visible (unmasked) patch embeddings. The sequence of visible patch embeddings which was augmented with positional embeddings and a class token, was passed through the stack of transformer encoder blocks, comprising multi-head self-attention and feed-forward layers. The encoder output a latent representation of size N × D, where N denotes the number of visible patches.

**MAE Decoder:** The decoder received the latent embeddings corresponding to visible patches, along with mask tokens representing the missing patches and their associated positional embeddings. The full token sequence was processed through transformer decoder blocks to predict the representations of the masked patches. A linear projection layer mapped the reconstructed embeddings back to pixel space to generate a full-resolution SNI reconstruction. The decoder was used exclusively during the pretraining stage for SNI reconstruction. The MAE model was optimized using the mean squared error (MSE) loss computed between the reconstructed and original SNIs in pixel space. The decoder was used solely during the self-supervised pretraining stage. After pretraining, the decoder was discarded and only the pretrained encoder was retained to provide feature representations for subsequent supervised tasks within the proposed pipeline.

## Model-agnostic knowledge augmentation-based classification module

After the pre-training stage of the self-supervised feature representation learning module, the encoder of the MAE model was finetuned for the SNI classification task using a small subset of labelled SNIs. To demonstrate the ability of our pipeline to perform under label scarce settings, we only used 10 labelled SNIs per class in each dataset in the finetuning stage. Since we only have a limited labelled dataset, we propose an annotation efficient classification method as demonstrated in Figure 1 Module (c). Our method augments the feature representations learned by pre-trained encoder of the MAE model using the prior knowledge based features from SNIs. To flexibly balance their respective contributions, we introduce a weighted knowledge augmentation mechanism that learns the weights assigned to automated features and manual features during model training.

**Prior knowledge based features:** We extracted a set of prior knowledge based features from the SNIs which were found relevant when analysing the TDP-43 localization pathology in microscopy images [29]. These features included the area of the nucleus, area of the surrounding of the nucleus (cytoplasm), average TDP-43 intensity inside the nucleus area, average TDP-43 intensity in the surrounding area of the nucleus, and ratio of TDP-43 in the surrounding area of the nucleus. The ratio of TDP-43 in the surrounding area of the nucleus is calculated using equation (1).

$$Ratio\ of\ TDP43\ in\ surrounding = \frac{Avg\ TDP43\ intensity\ in\ surrounding}{(Avg\ TDP43\ intensity\ in\ nucleus + Avg\ TDP43\ intensity\ in\ surrounding)} \quad (1)$$

For the cytomorphology datasets, we extracted 51 handcrafted features previously shown to be informative [28]. These included three nucleus shape descriptors, convexity, circularity, and solidity, alongside 48 colour features derived from multiple colour spaces (RGB, HSV, LAB, and YCrCb).

**Weighted Knowledge Augmentation:** The automated feature representations of the SNIs were obtained using the pre-trained encoder of the MAE model. The prior knowledge based features were fed through a linear layer to obtain the manual feature embedding. We introduce a learnable weighted concatenation mechanism as defined by equation (2):

$$Final\ Embedding\ of\ SNI = (w_{auto} \times Automated\ Feature\ Embedding) \oplus (w_{manual} \times Manual\ Feature\ Embedding) \quad (2)$$

where $\oplus$ and × indicate concatenation and multiplication respectively. Automated feature embedding refers to the feature representations obtained from the pre-trained encoder. The parameters $w_{auto}\ and\ w_{manual}$ were implemented as learnable scalar weights corresponding to automated feature embedding and manual feature embedding respectively, each initialized to 0.5 at the beginning of the training and summed to 1. Our classification module was trained end-to-end using the cross entropy loss between the predicted label and the ground truth label. This formulation allowed the model to automatically learn and adjust the contribution of each component during optimization, enabling adaptive knowledge augmentation without the requirement of manual tuning. Additionally, the automated feature embedding can be obtained from any ML model and can be integrated with prior knowledge based features using the proposed knowledge augmentation method.

**Classification of high green intensity SNIs:** As described in Results section, in datasets 2 and 3, a subset of SNIs exhibited abnormally elevated green channel intensity, necessitating their temporary treatment as a separate class ($SNI_{green}$). To prevent these samples from degrading downstream classification performance, we first isolated them as a provisional third category and subsequently reassigned them using an unsupervised clustering method. Specifically, feature embeddings were extracted from the pretrained MAE encoder and projected into a lower dimensional space using principal component analysis (PCA) [53]. Agglomerative clustering with cosine affinity and average linkage was then applied to the PCA transformed embeddings to partition the $SNI_{green}$ subset into two clusters corresponding to the underlying primary categories. In parallel, the pretrained MAE encoder was additionally fine-tuned to classify high green intensity SNIs, enabling the model to better capture the visual variations associated with this artefactual signal. This combined strategy ensured robust correction of intensity related anomalies while preserving the integrity of the final class assignments.

# Human Research Ethics

For dataset 1 and dataset 3, the study was conducted according to the National Statement on Ethical Research and approved by the University of Wollongong Human Research Ethics Committee (approval numbers 13/272 and 19/380). The iPSCs were cultured as previously reported in [54,55]. Briefly, the iPSCs were maintained in mTESR1 (StemCell Technologies, #85850) on matrigel (Biostrategy) coated 60 mm tissue culture dishes, kept in hypoxic conditions at 37 °C with 5 % CO2. Cells were passaged once every 5–7 days using 0.5 mM EDTA (Life Technologies, #AM9260G) in phosphate-buffered saline (PBS)−/− (Life Technologies, #14190250). Media changes were performed daily. Mycoplasma and STR profiling were performed routinely. For dataset 2, the project was approved by the University of Melbourne Human Research Ethics committee (ID: 1749960).

## Acknowledgements

This research was funded by a FightMND Impact grant (07) awarded to LO and SH and a US Department of Defense Therapeutic Idea Award (AL210047; W81XWH-22-1-0311) awarded to LO and SH. Parts of this research were supported by the Australian Research Council Centre of Excellence in Quantum Biotechnology (CE230100021) awarded to LO and a FightMND Drug Screening Program grant awarded to EQ. JB was supported by the Melbourne Research Scholarship.

## Competing interests

The authors declare no competing interests.